\documentclass[12pt,preprint2]{aastex}
\def\Msol{\thinspace\hbox{$\hbox{M}_{\odot}$}}

\def\a4{\hsize 17.0cm \vsize 25.cm}
\newcommand{\der}[2]  { \frac{{\rm d}#1}{{\rm d}#2} }

\newcommand{\dif}     {{\rm d}}

\shorttitle{superwinds \& supernebulae}
\shortauthors{Silich et al.}

\begin{document}

\title{On the extreme stationary outflows from super-star clusters: 
from superwinds to supernebulae and further massive star formation.}

\author{Guillermo  Tenorio-Tagle, Sergiy Silich, Ary Rodr\'\i{}guez-Gonz\'alez}
\affil{Instituto Nacional de Astrof\'\i sica Optica y
Electr\'onica, AP 51, 72000 Puebla, M\'exico; gtt@inaoep.mx}

\and

\author{Casiana Mu\~noz-Tu\~n\'on}
\affil{Instituto de Astrof\'{\i}sica de Canarias, E 38200 La
Laguna, Tenerife, Spain; cmt@ll.iac.es}

\begin{abstract}
Here we discuss the properties of star cluster winds in the supercritical, 
catastrophic cooling regime. We demonstrate that catastrophic cooling 
inhibits superwinds and after a rapid phase of accumulation of the
ejected material within the star-forming volume a new stationary
isothermal regime,  supported by the ionizing radiation from the
central cluster, is established. The expected appearance of this core/halo 
supernebula in the visible line regime and possible late evolutionary 
tracks for super-star cluster winds,  in the absence of ionizing radiation, 
are thoroughly discussed. 
\end{abstract}


\section{Introduction}

Within the volume occupied by a star cluster (SC), the energy injected 
by stellar winds and supernova explosions is fully thermalized via random 
interactions. This generates the large central overpressure that continuously 
accelerates the ejected gas and eventually blows it out of the star cluster 
volume to compose a superwind. In the adiabatic solution of Chevalier \& Clegg 
(1985; hereafter referred to as CC85; see also Canto, et al. 2000 and Raga 
et al. 2001) temperature and density present  almost homogeneous values 
within the central volume, whereas the expansion velocity grows almost 
linearly from 0 km s$^{-1}$ at the center, to the sound speed ($c$) at 
the cluster radius $r = R_{SC}$. There is then a rapid evolution as matter 
streams away from the star cluster and the wind parameters (velocity, 
density, temperature and pressure) soon approach their asymptotic values: 
$V_w \to $ $V_{A\infty} = (2 L_{SC}/{\dot M}_{SC})^{1/2} \sim 2c(R_{SC})$, 
$\rho_w \sim $ $r^{-2}$, $T_w \sim $ $r^{-4/3}$ and thus $P_w \sim 
r^{-10/3}$, where $L_{SC}$ and ${\dot M}_{SC}$ are energy and mass deposition
rates.
 
Recent results on the outflows expected from SCs have led us to realize that
the adiabatic steady wind solution proposed by CC85 and followers,
becomes inapplicable in the case of massive and concentrated clusters. 
Radiative cooling strongly modifies first the temperature distribution 
predicted for adiabatic stationary winds ($T_w \sim r^{-4/3}$) bringing 
suddenly and within  a small radius, the temperature down to 10$^4$ K, 
restricting then the X-ray emissivity of the winds to a volume much 
smaller than previously thought (see Silich et al. 2003 and 2004; hereafter 
referred to as Papers I and II). Also, as shown in Paper II,
for more energetic clusters,  strong radiative cooling  promotes the 
sudden leakage of thermal energy right within the star cluster volume itself,
and for the cases in which the radiative losses there exceed 30$\%$ of the 
stellar energy deposition rate, when cooling becomes catastrophic,   
then the stationary superwind solution is totally inhibited. In this latter 
case, the rapid drop in temperature within the SC volume, leads to a 
sudden drop in central pressure and particularly to a sudden drop in 
sound speed. This inhibits the fast acceleration predicted in the 
adiabatic solution as required by  the flow to reach its adiabatic terminal 
speed ($v_{A\infty} \sim 10^3$ km s$^{-1}$). The sudden drop in sound speed 
upsets also  the balance, demanded by the stationary solution, between 
the stellar mass input rate and the rate at which matter can flow away 
from the cluster, ie: $\dot M_{SC} = 4 \pi R_{SC}^2 \rho_{SC}(R_{SC}) 
c(R_{SC})$, and this inevitably leads to mass accumulation.

Here we show how the metallicity attained by  the ejected matter from
a coeval cluster, strongly affects the limits found in Paper II
bringing the energy input rate further down to much smaller values and
thus affecting even lower mass clusters. We also show how after the 
thermal stationary superwinds are inhibited, and after an inevitably 
short phase of matter accumulation within the SC volume, a second 
stationary solution can be found. This is only possible while the 
stellar UV flux ionizes the deposited matter to compose a dense supernebula  
with an isothermal ($T \sim 10^4$ K), slow ($v_{\infty} \sim 50 $ km s$^{-1}$),
stationary wind (see section 2). Finally, as the evolution continues and 
the stars producing the UV photon output evolve into supernovae, a third 
quasi-adiabatic stationary stage could become possible after a second phase 
of matter accumulation within the star cluster radius. This could in 
principle evolve into a massive, cold, neutral outflow, however, its maximum  
velocity is well below the escape speed and thus the ejected matter, 
unable to escape the gravitational potential of the star cluster, will 
sooner or latter recolapse to cause a second major burst of star formation.  
Our conclusions and the observational properties of the various 
evolutionary phases are given in section 3.

\section{The properties of massive star clusters}

Superstar clusters recently found by the Hubble Space Telescope in a large 
variety of starburst galaxies (for a review see Ho 1997 and the recent 
proceedings from "The formation and evolution of massive young star 
clusters",  Eds. Lamers et al, 2004) present a typical half-light radius 
$\sim$ 3 - 10 pc, and masses that range from several times $10^4$ M$_\odot$ 
to several times 10$^7$ M$_\odot$ (see Walcher et al. 2004). These are
now believed to be the unit of violent star formation in starburst galaxies.

The theoretical properties of massive bursts of star formation strongly 
depend on the assumed stellar evolutionary tracks. Synthesis models 
(e.g. Leitherer \& Heckman 1995, Leitherer et al. 1999) assume further a 
stellar IMF and an upper and lower mass limit for the star formation event. 
In this way one knows that a coeval burst of 10$^6$ M$_\odot$ with a 
Salpeter IMF, and stars between 100 M$_\odot$ and 1 M$_\odot$, would 
produce, through winds and supernovae (SNe), an almost constant mechanical 
energy input rate of $\sim 3 \times 10^{40}$ erg s$^{-1}$ for almost 50 Myr,
until the last star of 8 M$_\odot$ explodes as supernova. At the same time, 
the flux of ionizing radiation  is to remain constant at $\sim 10^{53}$ 
photons s$^{-1}$ for the first 3 Myr, to then steadily fall off 
approximately as $t^{-5}$. The above implies that the HII region phase 
will last for about 10 Myr, time during which the cluster photon flux 
will have decrease by more than two and a half orders of magnitude with 
respect to its initial value. The above values scale linearly, throughout 
the evolution,  with the mass of the assumed cluster and thus the HII region 
phase is in all coeval cases four to five times shorter than the supernova 
phase.

When dealing with the outflows generated by star clusters, another important 
intrinsic property is the metallicity of their ejected matter. This is a  
strongly varying function of time, bound by the yields from massive stars 
and their evolution time. Thus, once the cluster IMF and the stellar mass 
limits are defined, the resultant metallicity is an invariant curve, 
independent of the cluster mass. Here we consider  coeval clusters
with a Salpeter IMF, and stars between 100 M$_\odot$ and 1 M$_\odot$, as 
well as the evolutionary tracks with rotation of Meynet \& Maeder (2002)  
and an instantaneous mixing of the recently processed metals with the 
stellar envelopes of the progenitors (see Silich et al. 2001 and 
Tenorio-Tagle et al. 2003, for an explicit description of the calculations). 
This leads to metallicity values (using oxygen as tracer) that rapidly 
reaches 14 Z$_\odot$ (see Figure 1), and although steadily decaying afterwards,
the metallicity remains above solar values for a good deal of the
evolution (for more than 20 Myr), to then fall to the original metallicity 
of the parental cloud. One of the main effects of an enhanced metallicity 
of the ejecta is to boost its radiative cooling and in such a case, 
massive clusters may inevitably enter into the catastrophic cooling 
regime, envisaged in Paper II, to then find their stationary superwinds 
totally inhibited. 
%---------------------------------------------------------------
\begin{figure}[t]
\plotone{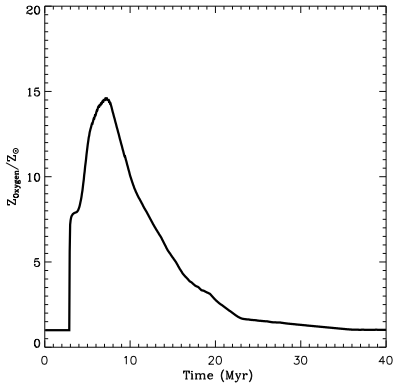}
\caption{The metallicity of the matter ejected by coeval bursts of star 
formation. The metallicity (in solar units) of the matter reinserted into 
the ISM by SCs is plotted as a function of the evolution time. The curve 
is derived from the  metal yields of Meynet \& Maeder (2002) stellar 
evolution models with stellar rotation, using oxygen as a tracer. The 
estimate assumes a Salpeter IMF and 100 M$_\odot$ and 1 M$_\odot$
upper and lower mass limit.  The model also assumes an instantaneous 
mixing between the newly processed metals and the envelopes of the
progenitors. Note that under these assumptions the curve is independent 
of the cluster mass.}
\label{fig1}
\end{figure}
%---------------------------------------------------------------
Relevant also is to note that at the end of the SN phase, the mass reinserted 
into the interstellar medium through winds and SNe amounts  to about 40$\%$ 
of the original mass in stars (see Leitherer \& Heckman 1995, their 
figure 53). Here we address the fate of such a vast amount of high 
metallicity matter.

\subsection{The limits between superwinds and supernebulae}

Figure 2 shows results from our self-consistent radiative code (see Paper II) 
indicating the limiting energy, as a function of the size of the clusters, 
at which radiative cooling becomes catastrophic for matter with metallicities 
similar to those expected for the gas emanating from massive stellar clusters. 
As pointed out in Paper II there are three different kinds of solutions: low 
energy input rates, produced by low-mass clusters, appear far from the 
threshold line that separates the catastrophic cooling regime from the 
stationary superwinds, and thus are to generate stationary quasi-adiabatic 
winds similar to those proposed by CC85. More energetic (or more massive 
clusters), as their energy input rate  approaches the threshold line will 
produce strongly radiative stationary superwinds with considerably smaller 
X-ray emitting volumes. Finally, cases that lie above  the threshold line 
in the catastrophic cooling zone, will radiate a large fraction of the 
energy input rate within the star cluster volume and thus will be unable 
to generate a stationary superwind.
%---------------------------------------------------------------
\begin{figure}[htbp]
\plotone{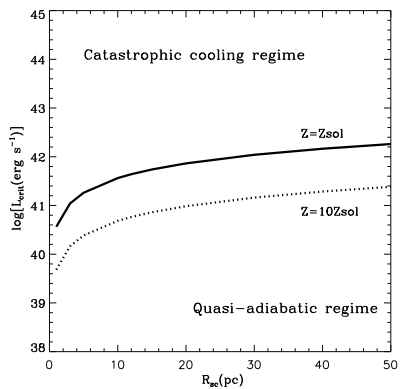}
\caption{The threshold energy input rate. The limiting energy input rate 
above which catastrophic cooling inhibits the thermal stationary superwind 
solution for SCs with $V_{A\infty} = 1000$ km s$^{-1}$, as a function of 
the size of the clusters and for two different abundances (solar and ten 
times solar) of the matter injected by the SCs.}
\label{fig2}
\end{figure}
%---------------------------------------------------------------

Figure 3 incorporates the run of the metallicity of the matter injected by 
winds and SNe as a function of time (Figure 1) to the results from our 
self-consistent radiative solution (Figure 2) and plots the resultant 
location of the threshold line for clusters with a $R_{SC}$ = 3 pc 
(solid line) and 10 pc (dashed line), the typical range of sizes 
of superstar clusters, as a function of time. As mentioned above, 
clusters of any given mass will evolve at an almost constant mechanical 
luminosity, crossing the diagram from left to right (as the $10^6$ M$_\odot$ 
cluster shown in the figure as a solid arrow). The figure confirms that 
low mass clusters  will produce quasi-adiabatic stationary winds. 
However, clusters with a mass in the range  3 $\times 10^5$ M$_\odot$ to 
3 $\times 10^6$ M$_\odot$ would cross twice the threshold line during their 
evolution and thus will have their thermal stationary winds fully inhibited 
for a good fraction of their evolution. More massive and compact clusters, 
those injecting more than say, 10$^{42}$ erg s$^{-1}$ (see Figure 3), 
will be unable to develop the superwind outflows during their evolution.

All of the latter clusters, unable to rid themselves from the continuous 
input of matter from their stellar sources, are to cause a rapid 
accumulation of the cold ($T \sim 10^4$ K) recombined ejecta, now suddenly 
exposed to the UV photon flux from the most massive stars. As shown below, 
this leads to a dense and compact, high metallicity  supernebula, able to 
establish an isothermal stationary HII region wind.  
%---------------------------------------------------------------
\begin{figure}[ht]
\plotone{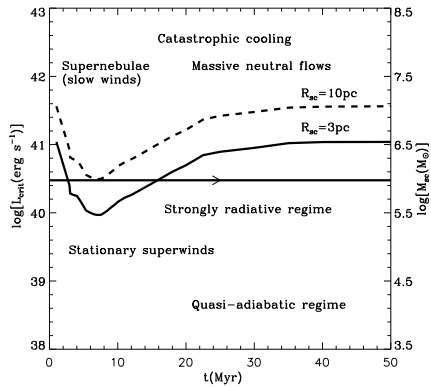}
\caption{The evolution of the threshold energy input rate. The continuous 
changes in the metallicity of the ejecta (Figure 1) shift the location of 
the threshold line. The figure shows the run of the threshold energy input 
rate for massive clusters with 3~pc (solid line) and 10~pc (dashed line) 
radii, respectively. The horizontal solid line across the diagram indicates 
the energy deposition rate of a 10$^6$ M$_\odot$ cluster. The scale for 
different mass clusters is indicated on the right-hand axis. The areas 
occupied by the various stationary solutions described in this study are 
also indicated in the figure.}
\label{fig3}
\end{figure}
%--------------------------------------------------------------- 

\subsection{The evolution of supernebulae}

For clusters that enter the catastrophic cooling regime,  matter accumulation 
within the star cluster volume, would rapidly lead to larger densities, while 
the stellar UV photon flux independently will sustain  the temperature at 
$T \sim 10^4$ by photoionization.  

The density $\rho_{SC}$ will increase until the stationary condition 
$\dot M_{SC} = 4 \pi R_{SC}^2$ $\rho_{SC}(R_{SC}) c(R_{SC})$ (where now 
$c(R_{SC})= c_{HII}$ $\sim 10$ km s$^{-1}$) is once again fulfilled. 
Given the drastic drop in temperature of the ejecta that results from 
thermalization followed by catastrophic cooling (say from 10$^7$ K to
10$^4$ K), the density $\rho_{SC}$ will have to increase by one and a half 
orders of magnitude to compensate the drop in sound speed ($\sim T^{0.5}$) 
and then meet the isothermal stationary condition. Mass accumulation will 
last for a short time ($\tau$)
%--------------------------------------------------------------- 
\begin{equation}
      \label{eq.1} 
\tau = \frac{4 \pi}{3} \frac{(\rho_2 -\rho_1) R_{SC}^3 V^2_{A\infty}}{2 L_{SC}}
       \sim 10^5 \, yr.
\end{equation}  
%--------------------------------------------------------------- 

Once this happens, an isothermal, stationary, photoionized wind will 
begin to emanate from the star cluster surface. The isothermal steady 
state flow equations within the star cluster radius ($R \le R_{SC}$) are
%---------------------------------------------------------------
\begin{eqnarray}
     \label{eq.2a} 
      & & 
\der{u_w}{r}  = \frac{q_m}{3\rho_w} \frac{1 + 3 u^2_w/c^2_{HII}}
                {1  - u^2_w/c^2_{HII}} , 
      \\[0.2cm] & & 
\rho_w = \frac{q_m r}{3 u_w} , 
\end{eqnarray}
%-------------------------------------------------------------
and outside of the star cluster ($R > R_{SC}$),  
%---------------------------------------------------------------
\begin{eqnarray}
      \label{eq.2b}
      & & 
\der{u_w}{r}  = \frac{2 u_w}{r} \frac{c^2_{HII}/u^2_w}
                {1 - c^2_{HII}/u^2_w} ,
      \\[0.2cm] & & 
\rho_w = \frac{{\dot M}_{sc}}{4 \pi u_w r^2} ,
\end{eqnarray}
%-------------------------------------------------------------
where $q_m$ is the mass deposition rate per unit volume inside a star
cluster, $c_{HII} = (k T_w / \mu_p)^{1/2}$ is the isothermal sound
speed, $\mu_p = 14/23 m_H$ is the mean mass per particle, 
$k$ is the Boltzmann constant, and $\gamma$ is the ratio of the specific
heats. The wind central temperature is supported by photoionization at a
$T_c \sim 10^4$~K level, and the wind central density can be found by
iterations from the condition that the isothermal sonic point
$(u_w = c_{HII})$ acquires its proper position at the star cluster
surface, $R_{sonic} = R_{sc}$ (for more details see paper II). The
isothermal wind temperature, density and velocity distributions are 
presented in Figure 4a-c (dashed lines).
%---------------------------------------------------------------
\begin{figure}[htbp]
\plotone{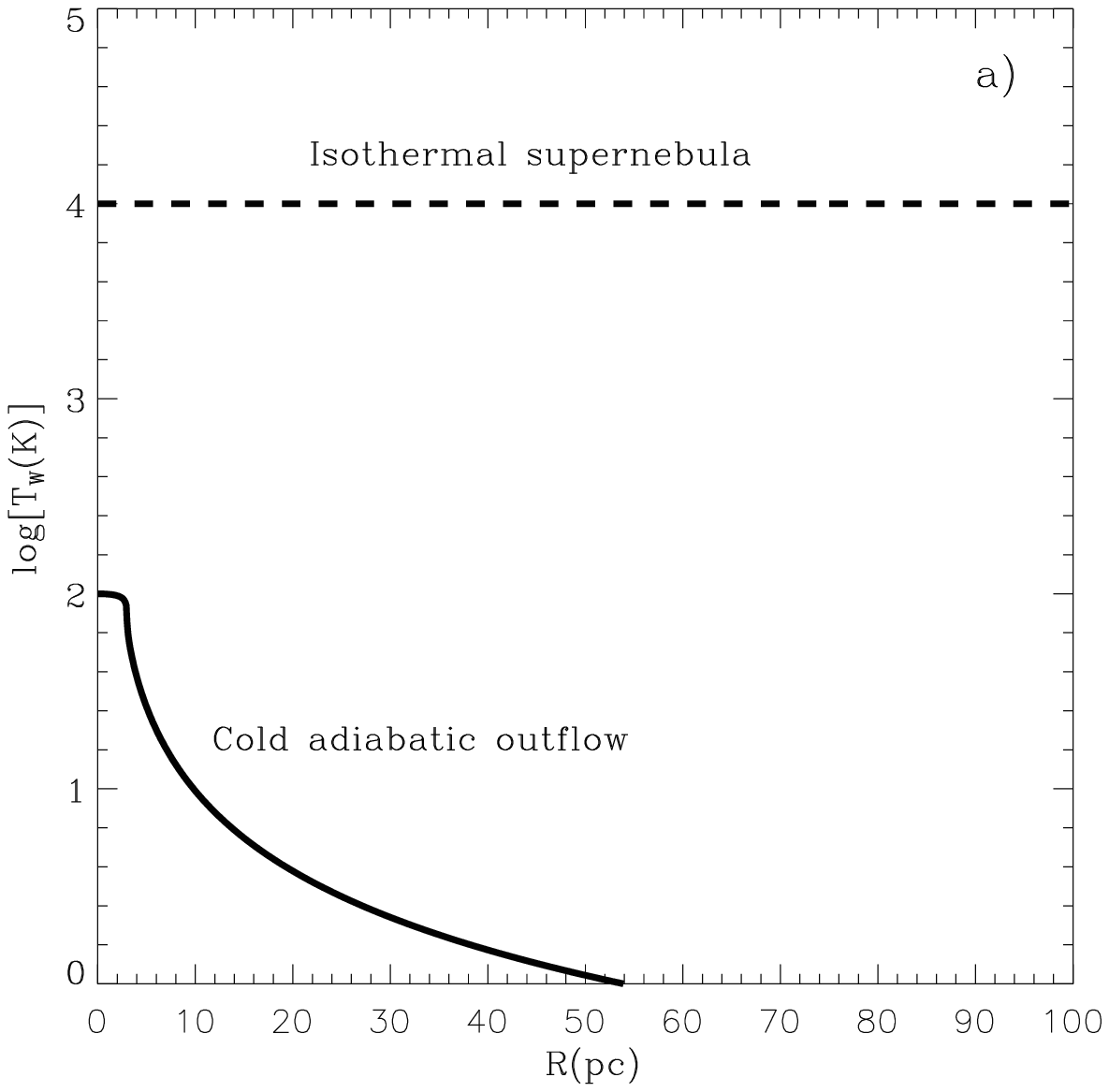}
\plotone{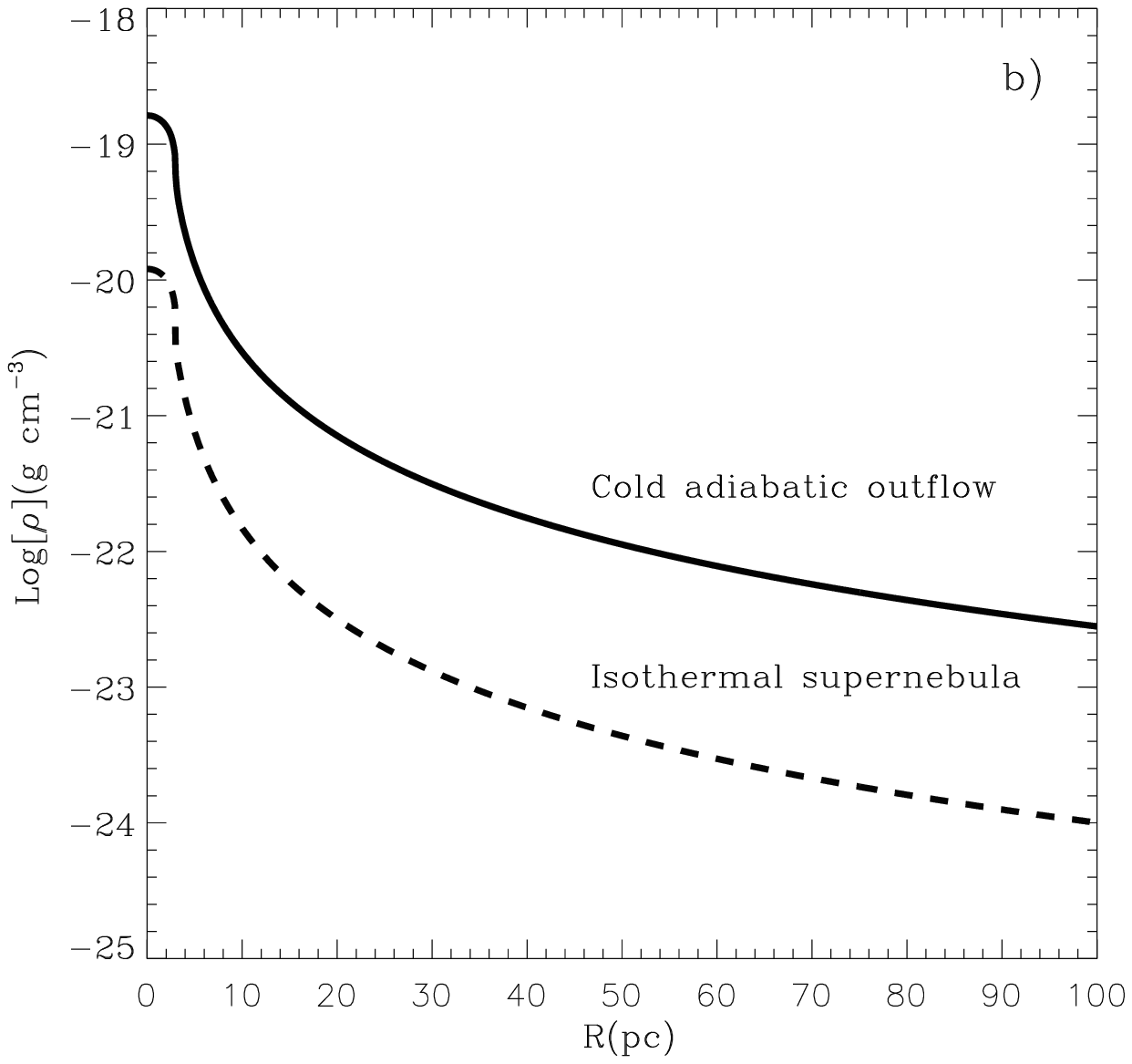}
\end{figure}
\begin{figure}[htbp]
\plotone{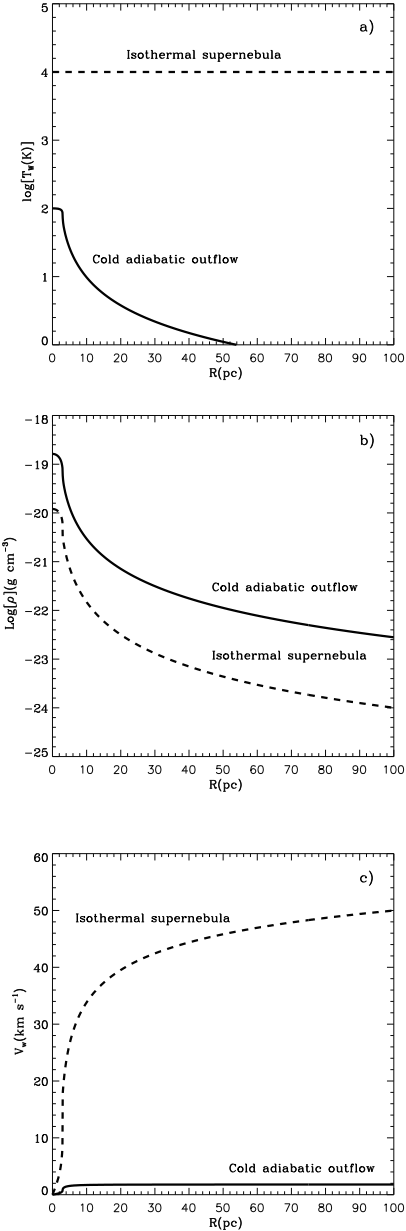}
\caption{Supernebulae and massive slow neutral outflows. The stationary 
inner structure (temperature, density and velocity, (a-c, respectively) 
as a function of $R$ (in pc), adopted by the outflow from a 10$^6$ 
M$_\odot$ compact (3 pc radius) cluster is shown for the supernebula 
(dashed lines) and massive slow neutral outflow phase (solid lines).}
\label{fig4}
\end{figure}
%---------------------------------------------------------------

The resultant core/halo supernebula density distribution (see Figure 4) can 
be approximated by a constant value $n_c$ within  $R_{SC}$ and by a power 
law  ($r^{-2}$) for $r > R_{SC}$. In this case, the number of recombinations 
per unit time throughout the supernebula volume is:
%---------------------------------------------------------------------- 
%\begin{equation}
\begin{eqnarray}
      & & \label{eq.3} \nonumber
\der{N_{rec}}{t} = 
%      \\[0.2cm]     \label{eq.3}
4 \pi \beta \left[
    \int_{0}^{R_{SC}} n^2(r) r^2 \dif r + \right.
      \\[0.2cm] & & \nonumber
   \left.   \int_{R_{SC}}^{R_{St}} n^2(r) r^2 \dif r \right] =
      \\[0.2cm] & & 
    \frac{16 \pi}{3} \beta n^2_c R_{SC}^3 \left[1 - \frac{3}{4} 
    \frac{R_{SC}}{R_{St}}\right] ,
\end{eqnarray} 
%\end{equation} 
%----------------------------------------------------------------------
where $R_{St}$ is the Str${\ddot o}$mgren radius and $\beta$ is the 
recombination coefficient to all but the ground level.
The Str${\ddot o}$mgren radius is then:
%---------------------------------------------------------------------- 
\begin{equation}
      \label{eq.4} 
R_{St} = \frac{3}{4} 
            \frac{R_{SC}}{1 - \frac{3 {\dot N}_{rec}}
            {16 \pi \beta n^2_c R^3_{SC}}} .
\end{equation} 
%----------------------------------------------------------------------
Equation (\ref{eq.4}) implies that the supernebula should be completely 
ionized by the star cluster UV radiation if the number of photons per 
unit time exceeds the critical value
%---------------------------------------------------------------------- 
\begin{equation}
      \label{eq.5} 
{\dot N}_{crit} = \frac{16 \pi}{3} \beta n^2_c R^3_{SC} . 
\end{equation} 
%----------------------------------------------------------------------
The bright, centrally concentrated high metallicity nebula and its moderate  
wind, causing a stationary core/halo structure, are to be maintained for as 
long as they remain fully ionized. Note however that as the flux of UV 
photons from an instantaneous, or coeval, burst of star formation declines 
drastically after $\sim 3 $~Myr (see Leitherer \& Heckman 1995) it would  
eventually fall below the critical value. The continuously reduced  number 
of ionizing photons, unable to balance the large number of recombinations 
in the ionized volume, forces the ionization front to recede supersonically 
towards the stars (see Beltrametti et al. 1981). From equation
(\ref{eq.4}) one can 
show that it will reach $R_{SC}$ when the number of ionizing photons drops 
by a factor of four below the critical value. At this time the central HII 
region will begin to loose its identity and uniform structure, to evolve 
finally into a collection of individual ultracompact HII regions around the 
most massive members left within  the coeval star cluster. 

Taking into account that the  supernebula central density $\rho_c = 
{\dot M}_{SC} / 4 \pi R^2_{SC} c_{HII}$ and that the star cluster 
mechanical luminosity $L_{SC} = \frac{1}{2} {\dot M}_{SC} V_{A, \infty}^2$, 
where $V_{A, \infty}$ is the adiabatic wind terminal speed in the 
hypothetical absence of radiative cooling, one can rewrite equation 
(\ref{eq.5}) in the form:
%---------------------------------------------------------------------- 
%\begin{equation}
\begin{eqnarray}
      & & \label{eq.6}  \nonumber
{\dot N}_{crit} = \frac{4 \beta L^2_{SC}}{3 \pi \mu_a^2 R_{SC} 
V^6_{A, \infty}} \left(\frac{V_{A, \infty}}{c_{HII}}\right)^2 =
      \\[0.2cm] & & 
\frac{\beta \dot M_{SC}^2}{3 \pi \mu_a^2 R_{SC} c_{HII}^2} 
\end{eqnarray}    
%\end{equation}
%----------------------------------------------------------------------
where $\mu_a$ is the mean mass per atom ($\mu_a = 14/11 m_H$) and 
the supernebula isothermal sound speed $c_{HII} \approx 11.6$ km s$^{-1}$. 
In such a case, our standard $10^6$\Msol, $R_{SC} = 3$~pc cluster and 
its supernebula will require of an ${\dot N}_{crit} \approx 1.7 \times 
10^{52}$~s$^{-1}$. This value is almost an order of magnitude below 
the maximum number of UV photons emitted initially by our example 
10$^6$ M$_\odot$ star cluster. The  UV flux from the aging cluster 
would however reach the critical value after 4 Myr and 5 Myr,
depending if one assumes solar or 0.1 solar as the original 
cluster metallicity  (see Leitherer \& Heckman, 1995; their Figure 37), 
and the UV radiation a couple of Myr after this will not suffice to sustain 
even the star cluster volume fully ionized. This brings  the supernebula 
phase to an end, while restricting the ionized volume to ultracompact HII 
regions around the most massive sources left in the cluster. Without the 
sufficient UV photons to maintain the SC volume fully ionized and at a 
temperature $\sim 10^4$ K, the stationary isothermal wind condition is 
inhibited and a new phase of matter accumulation will start. 

Figure 5 displays the H$_\alpha$ luminosity of our 10$^6$ M$_\odot$ cluster 
that injects 3 $\times 10^{40}$ erg s$^{-1}$ within a $R_{SC} = $ 3 pc and 
thus initially, while the metallicity remains below Z$_\odot$, it produces 
a strongly radiative stationary superwind (follow the horizontal line in  
Figure 3). The stationary strongly radiative solution leads to a temperature 
$\sim 5 \times 10^5$ K at a distance of 17.5 pc and thus at this radius the 
streaming wind matter recombines and is exposed to the stellar UV photon 
output. Photoionization causes in such a case, a thin stationary shell 
with the matter that continuously streams with large speeds ($\sim 10^3$ 
km s$^{-1}$) across the recombining radius and leads thus to a top-hat 
line profile (Rodr\'\i{}guez-Gonz\'alez et al. 2004; Owocki \& Cohen,
2001; Dessart \& Owocki, 2002) shown by dotted line in Figure 5. This
is very different to the calculated Gaussian line profile produced
during the supernebular phase, which displays a linewidth of the order 
of 100 km s$^{-1}$ (FWZI) and, given the large densities, has an
intensity several orders of magnitude larger than that of the top-hat 
line profile produced during the initial superwind stage. The total 
integrated intensity over the line profiles is: 5.3 $\times 10^{40}$ erg 
s$^{-1}$ for the supernebula stage and 6.9 $\times 10^{34}$  erg
s$^{-1}$ for the strongly radiative superwind.
%---------------------------------------------------------------
\begin{figure}[t]
\plotone{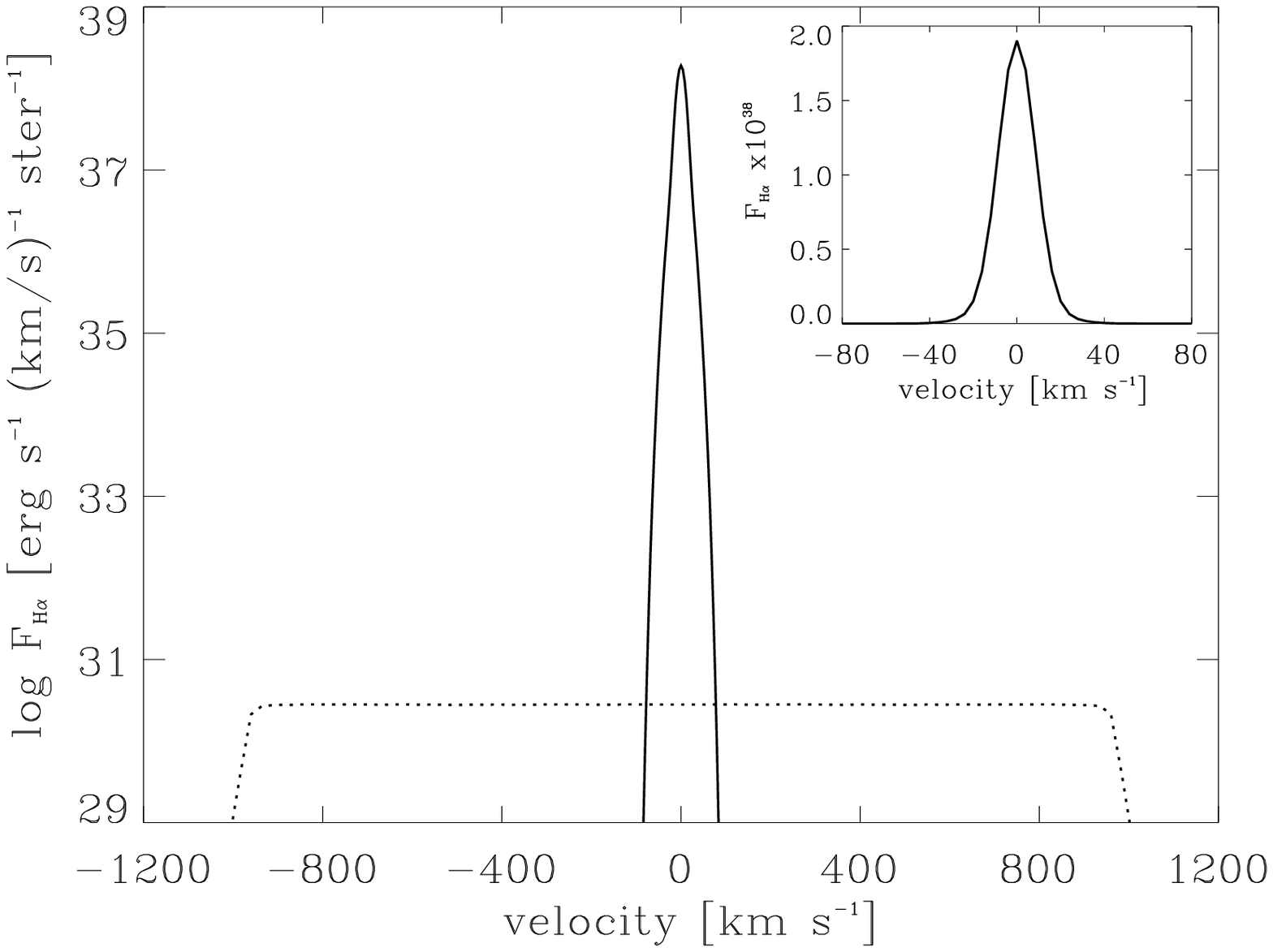}
\caption{The signature of stationary strongly radiative superwinds and 
supernebulae. The H$_\alpha$ line profiles produced by  the outflow from 
a compact (3 pc radius), 10$^6$ M$_\odot$ star cluster, at different 
stages of its evolution. The dotted line represents the low intensity, 
flat-top H${_\alpha}$ profile  calculated for a strongly radiative superwind 
stage, before the metallicity of the outflowing gas exceeds Z$_\odot$. 
In such a case strong radiative cooling brings the superwind temperature 
down to $T \sim 10^4$ K at a distance of 17.5 pc allowing the stellar UV 
photon output to ionize the rapidly moving stream, while causing a broad 
flat-top line profile. This is to be compared with the line profile 
calculated for the supernebula phase (solid lines). The total integrated 
intensity over the line profiles is: $6.9 \times 10^{34}$ erg s$^{-1}$ 
and $5.3 \times 10^{40}$  erg s$^{-1}$, respectively.}
\label{fig5}
\end{figure}
%---------------------------------------------------------------

\subsection{Gravitationally bound massive neutral outflows.}

The lack of sufficient UV photons to keep the supernebula phase at work, 
inevitably leads to a further accumulation of the ejected matter within 
the star cluster volume, until the density there raises by another order 
of magnitude and compensates in this way the drop in temperature 
(say, from 10$^4$ K to 100 K) and a stationary flow can once again be 
established. This time however, the quasi-adiabatic flow is very massive 
and very slow ($v_{\infty} \sim$ 2 km s$^{-1} \sim 2 c_{H_2}$).
Figure 4 a-c (solid lines) display the properties (the run of temperature, 
density and velocity as a function of distance to the star cluster center)
of such stationary flow. 

Figure 4 compares the isothermal ($T \sim 10^4$ K) high metallicity 
supernebula outflow with the late massive adiabatic neutral flow.
In both cases the density drops as $r^{-2}$ and thanks to the  continuous  
energy input rate through photoionization the  supernebula is able to expand 
with increasingly larger velocities (up to 50 km s$^{-1}$). 
The adiabatic dense outflow case reaches instead a maximum outflow 
velocity of only $v \sim 2 c \sim$ 2 km s$^{-1}$. This maximum speed is well 
below the escape speed ($v_{esc} = (2GM_{SC}/R_{SC})^{0.5}$)
from the central cluster and thus is to evolve instead into another phase of 
matter accumulation, to eventually recolapse into a new generation of stars; 
subject of a forthcoming communication.

\section{Feedback and the observational properties of the superstar cluster 
         outflows.}

Contrary to the predictions from the adiabatic solution of Chevalier \& 
Clegg 1985, we have shown here that massive and concentrated star clusters 
do not necessarily blow a strong stationary superwind throughout their 
evolution. The high metallicity of the matter deposited through winds and 
supernovae, enhances radiative cooling within the star cluster volume, 
and  leads instead to two other possible types of stationary flows emanating 
from massive SC. These have been termed here  supernebula and massive slow 
neutral outflows. The high metallicity supernebulae with a distinct core/halo 
density distribution, expanding with up to 50 km s$^{-1}$, can only be 
maintained in presence of an ample supply of stellar UV photons. These 
should be able to keep, at least, the central SC volume fully ionized. 
We have shown that this phase is thus bound to the first 10 Myr of the 
cluster evolution. 

On the other hand, the slow massive neutral flows, that follow the 
supernebula phase, are to last in the case of very massive clusters 
(M$_*$ $\geq 3 \times 10^6$ M$_\odot$), until the end of the supernova 
phase ($\sim$ 40 - 50 Myr). The maximum speed of the neutral outflows is 
however much slower than the escape speed from the massive clusters
and thus are to be inhibited by the potential of the clusters, leading 
instead to a further accumulation of matter which may eventually collapse 
and lead to a new stellar generation.

Very massive clusters, above the threshold line, will enter from the start 
of their evolution the supernebula phase, and will never cause a superwind. 
On the other hand, lower mass clusters will never experience 
either the supernebula nor the massive neutral stationary outflows and will 
establish stationary quasi-adiabatic superwinds throughout their evolution. 
Note however, that coeval clusters with a mass in the range 3 $\times 10^5$ 
M$_\odot$ to 3 $\times 10^6$ M$_\odot$ and a $R_{SC}$ = 3 pc (see Figure 3) 
will start their evolution causing the development of a stationary strongly 
radiative superwind, to then enter into the supernebulae phase and finally, 
after 8 - 20 Myr, respectively, re-enter once again into the stationary 
strongly radiative superwind regime. Clearly, this will lead to a very 
strong interaction between the new superwind and the matter left near the 
cluster during the supernebula or the massive and slow neutral outflow phases.

All superstar clusters will be, throughout their evolution, strong X-ray 
emitting sources. Clusters below the threshold line, that separates 
stationary superwinds and catastrophic cooling solutions (see Figure 3),
are to emit copiously in X-rays both within the SC central volume and in 
their winds. Clusters above the threshold line, whether experiencing the  
supernebula or  the  massive slow neutral outflow, will also emit in 
X-rays within their central volume, as catastrophic cooling radiates away 
the continuously replenished thermal energy. The X-ray properties 
of such clusters, compared to the X-ray emission from superbubbles will be 
the subject of a forthcoming communication.

Our gratitude to Leticia Carigi for her suggestions regarding the metallicity 
of the ejected matter. We thank our anonymous referee for his comments
and suggestions. This study has been supported by CONACYT - M\'exico, 
research grant 36132-E and the Spanish Consejo Superior de Investigaciones 
Cient\'\i{}ficas, grant AYA2001-3939.

%\newpage

%\onecolumn


%\begin{figure}
%\figcaption[ms15457_1.ps]
%{The dispersion relation $\omega (\eta )$ for $c_{sh}$ corresponding to 
%position of a 100 M$_{\odot}$ fragment. \label{fig1}}
%\end{figure}

\end{document}